\documentclass[sigconf]{acmart}

\copyrightyear{2024}
\acmYear{2024}
\setcopyright{rightsretained}
\acmConference[ICSE '24]{2024 IEEE/ACM 46th International Conference on Software Engineering}{April 14--20, 2024}{Lisbon, Portugal}
\acmBooktitle{2024 IEEE/ACM 46th International Conference on Software Engineering (ICSE '24), April 14--20, 2024, Lisbon, Portugal}\acmDOI{10.1145/3597503.3639144}
\acmISBN{979-8-4007-0217-4/24/04}
\usepackage{amsmath}
\usepackage[utf8]{inputenc} 
\usepackage[T1]{fontenc}    
\usepackage{booktabs}
\usepackage{bbding}
\usepackage{microtype}      
\usepackage{hyperref}
\usepackage{algorithmic}
\usepackage{graphicx}
\usepackage{textcomp}
\usepackage{xcolor}
\usepackage{multirow}
\usepackage{enumitem}
\usepackage{colortbl}
\usepackage{nicefrac}       
\usepackage{hhline}
\usepackage[skip=1pt,font=small,labelfont=bf]{caption}
\usepackage{pifont}
\usepackage{subcaption}
\usepackage{xspace}
\usepackage[normalem]{ulem}
\useunder{\uline}{\ul}{}
\usepackage{tcolorbox}
\def\BibTeX{{\rm B\kern-.05em{\sc i\kern-.025em b}\kern-.08em
    T\kern-.1667em\lower.7ex\hbox{E}\kern-.125emX}}
    
\usepackage[linesnumbered,ruled,vlined]{algorithm2e}
\SetKwInput{KwInput}{Input}                
\SetKwInput{KwOutput}{Output}              
\usepackage{multicol}

\usepackage{array}

\usepackage[linesnumbered,ruled,vlined]{algorithm2e}

\usepackage{xspace}
\makeatletter
\DeclareRobustCommand\bmvaOneDot{\futurelet\@let@token\bmv@onedotaux}
\def\bmv@onedotaux{\ifx\@let@token.,\else.\null\fi\xspace}
\DeclareRobustCommand\bmvaTwoDot{\futurelet\@let@token\bmv@twodotaux}
\def\bmv@twodotaux{\ifx\@let@token.,\else.,\null\fi\xspace}
\makeatother
\def\eg{\emph{e.g}\bmvaTwoDot} 
\def\ie{\emph{i.e}\bmvaTwoDot} 
\def\cf{\emph{cf}\bmvaOneDot}

\def\etal{\emph{et al}\bmvaOneDot}

\newcommand{\toolname}{\textit{REOM}\xspace}

\def\vec#1{{\boldsymbol{#1}}}

\newenvironment{RuleList*}[1][htb]
  {
   \begin{algorithm*}[#1]%
  }{\end{algorithm*}}
\usepackage{makecell}

\begin{document}

\title{Investigating White-Box Attacks for On-Device Models
}

\author{Mingyi Zhou}
\email{mingyi.zhou@monash.edu}
\affiliation{%
  \institution{Monash University}
  \city{Melbourne}
  \state{VIC}
  \country{Australia}
}
\author{Xiang Gao}
\email{xiang_gao@buaa.edu.cn}
\affiliation{%
  \institution{Beihang University}
  \city{Beijing}
  \country{China}
}
\author{Jing Wu}
\email{jing.wu1@monash.edu}
\affiliation{%
  \institution{Monash University}
  \city{Melbourne}
  \state{VIC}
  \country{Australia}
}

\author{Kui Liu}
\email{brucekuiliu@gmail.com}
\affiliation{%
  \institution{Huawei Software Engineering Application Technology Lab}
  \country{China}
}

\author{Hailong Sun}
\email{sunhl@buaa.edu.cn}
\affiliation{%
  \institution{Beihang University}
  \city{Beijing}
  \country{China}
}

\author{Li Li}\authornote{Dr. Li Li was a senior lecturer at Monash. He supervised this project for the whole period. Corresponding authors: Li Li.}
\email{lilicoding@ieee.org}
\affiliation{%
  \institution{Beihang University, Beijing}
  \city{Yunnan Key Laboratory of Software Engineering}
  \country{China}
}

\begin{abstract}

Numerous mobile apps have leveraged deep learning capabilities.
However, on-device models are vulnerable to attacks as they can be easily extracted from their corresponding mobile apps. 
Although the structure and parameters information of these models can be accessed, existing on-device attacking approaches only generate black-box attacks (\ie indirect white-box attacks), which are less effective and efficient than white-box strategies.
This is because mobile deep learning (DL) frameworks like TensorFlow Lite (TFLite) do not support gradient computing (referred to as non-debuggable models), which is necessary for white-box attacking algorithms. 
Thus, we argue that existing findings may underestimate the harmfulness of on-device attacks.
To validate this, we systematically analyze the difficulties of transforming the on-device model to its debuggable version and propose a Reverse Engineering framework for On-device Models (\toolname), which automatically reverses the compiled on-device TFLite model to its debuggable version, enabling attackers to launch white-box attacks.
Our empirical results show that our approach is effective in achieving automated transformation (\ie 92.6\%) among 244 TFLite models.
Compared with previous attacks using surrogate models, \toolname enables attackers to achieve higher attack success rates (10.23\%$\to$89.03\%) with a hundred times smaller attack perturbations (1.0$\to$0.01). 
Our findings emphasize the need for developers to carefully consider their model deployment strategies, and use white-box methods to evaluate the vulnerability of on-device models.
Our \textbf{artifacts} \footnote{\href{https://github.com/AnonymousAuthor000/code157}{https://github.com/zhoumingyi/REOM}} are available.

\end{abstract}

\maketitle


\section{Introduction}

The number of mobile devices worldwide is continuously growing. The capabilities of those devices also keep increasing, i.e., with powerful Central Processing Units (CPUs) and a large amount of memory, making them suitable for running deep learning (DL) models. Indeed, mobile devices have now become an ideal platform for deploying the DL model. Many intelligent applications have already been deployed on mobile devices~\cite{xu2019first} and have already benefited millions of users. 
Though DL models can also be deployed on a cloud platform, data transmission between a mobile device and the cloud may compromise user privacy. Indeed, to achieve high-level security,  users' personal data should not be sent outside the device.
This could be the reason why more and more DL models are deployed on the device, which has been advertised as one of the most important features by the newly emerged OpenHarmony mobile system~\cite{li2023software}), and the corresponding models are often referred to as on-device models.

\begin{figure}[ht]
  \begin{center}
    \includegraphics[width=\linewidth]{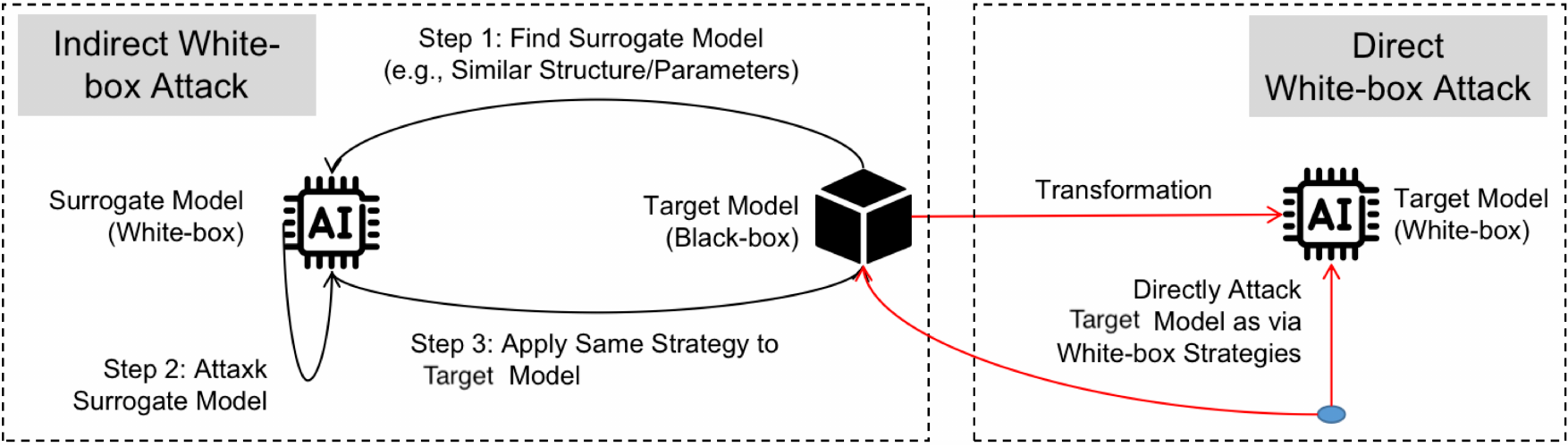}
  \end{center}
  \caption{The typical scenarios of evaluating the vulnerability of on-device DL models.}
  \label{fig:attack_motivation}
\end{figure}

Unfortunately, such on-device models are directly presented on mobile devices, giving attackers a lot of opportunities to exploit since it is relatively easy to unpack mobile apps to locate the physical models.
As a result, on-device models are facing more and more serious security threats.
Although on-device models are released to users as black-box ones for preventing potential attacks because attackers cannot obtain gradient information\footnote{Gradient information is considered crucial to implement effective white-box attacks.} from on-device models (referred to as \textbf{non-debuggable models}), they do not fulfill such a purpose in practice.
Indeed, as illustrated in Figure~\ref{fig:attack_motivation}, attackers still find ways to attack black-box models without accessing their gradient information, e.g., via the so-called \textbf{transferable attacks} (referred to as \textbf{indirect white-box attack}).
They achieve this by first, for target models, identifying debuggable surrogate models that are available to generate attacks.
They then exploit surrogate models through white-box strategies.
Once the strategies satisfy the attackers' needs, they apply the same strategies to attack on-device models.

In fact, many vulnerabilities of on-device models have already been discovered by our fellow researchers in recent years.
For example, Huang et al.~\cite{huang2021robustness,huang2022smart} propose to achieve the purpose by parsing features of the on-device model to find a surrogate model from the web, which could then be used to launch transferable attacks on mobile models.
Cao \etal~\cite{cao2021towards} also use surrogate models for attacking mobile models under the black-box setting, albeit by obtaining information from mobile models via querying their outputs, and then training a surrogate model using such information. 

Unfortunately, the performance of these approaches is highly dependent on the similarity between the surrogate model and the target models.
It is often difficult to find an ideal surrogate model that is highly similar to the target, thus affecting the effectiveness of attacks.
Since on-device models are directly hosted on devices, we manually look into those models and observe that such on-device models still keep the model's architecture and weights information but cannot be directly accessed.
We are, therefore, wondering if it is possible to extract this information so as to allow for evaluating vulnerabilities of on-device models as if they are white-box ones, without generating surrogate models (\cf the right part in Figure~\ref{fig:attack_motivation}).
If that is possible, current evaluation methods underestimate the threats of on-device models as direct white-box attacks are much more effective than black-box attacks (\ie indirect white-box attacks)~\cite{rosenberg2021adversarial}.
To this end, in this work, we propose a method \toolname to explore the following research question:

\textbf{RQ: Can on-device Models Be Directly Attacked via White-box Strategies?}

Borrowing the idea of reverse engineering the software artifact, which is often considered a black box as analysts cannot directly access the code, we start by checking if it is possible to reverse engineering on-device models.
Our exploitation reveals that it is possible to obtain a white-box version of the target model by first transforming it to an Open Neural Network Exchange (ONNX) model
and then transforming it back to debuggable AI models (thanks to the transparency of ONNX).

Towards answering the aforementioned research question, we start by verifying this hypothesis through a preliminary study.
Specifically, we focus on transforming TensorFlow-Lite (TFLite in short) models, the most popular on-device models, to ONNX and then PyTorch models, which is the most popular debuggable model format.
With 244 TFLite models extracted from 173,905 Google Play apps released in 2021, preliminary experimental results show that such a process is not able to achieve our purpose, i.e., transforming black-box on-device models with white-box versions.
Indeed, over 90\% of the models cannot be successfully transformed.

We then go one step deeper to understand why the majority of models cannot be transformed by looking into the error messages.
Our manual investigation reveals three types of errors in the above model transformation study, namely \texttt{Compatibility Errors}, \texttt{Not Implemented Errors}, and \texttt{Input Type Errors}.
To this end, we propose to complement the aforementioned process by automatically correcting those errors.
In particular, we present to the community a model transformation framework called \toolname, which includes dedicated strategies in three modification modules to correct the aforementioned three types of errors, respectively.
Experimental results show that our approach is effective by increasing the success rate from 6.6\% to 92.6\%, with the aforementioned 244 on-device TFLite models.
We further demonstrate that the transformed debuggable models and the original on-device model are indeed very similar, with a normalized $\ell_2$ output distance less than 0.001 in most cases.
Moreover, we also experimentally show that the transformed models can indeed support stronger attacks. 
Compared with previous methods of generating attacks using surrogate models, attackers can achieve higher attack success rates (10.23\%$\to$89.03\%) with a hundred times smaller attack perturbations (1.0$\to$0.01) based on our proposed tool. 

The main contributions of this paper are shown as follows:
\vspace{-0.5em}

\begin{itemize}[leftmargin=*]
  \item We propose a complete Reverse Engineering framework for On-device Models (\toolname) to convert the compiled on-device models to their corresponding debuggable version.
  \item \toolname can transform the model automatically, which presents the potential to be an essential tool to develop methods for testing the reliability of on-device models.
  \item Our paper shows attackers can achieve comparable on-devices attack performance with the white-box setting. The current model deployment strategy is at serious risk.
  \item We provide solutions to defense against reverse engineering based on our observation.
\end{itemize}

\section{Background and Related Work}
\label{sec:bg}

We now provide the necessary background about on-device DL models, DL model attack,s and the ONNX project.

\subsection{On-device DL Models}
\paragraph{\textbf{DL frameworks}:}
The open-source community has developed many well-known open-source frameworks for DL tasks such as TensorFlow~\cite{tensorflow2015_whitepaper}, Theano~\cite{al2016theano}, Caffe~\cite{jia2014caffe}, Keras~\cite{chollet2018keras}, and PyTorch~\cite{paszke2019pytorch}. These frameworks dominate the development of DL models and set standards for them~\cite{dilhara2021understanding}. 
PyTorch is one of the latest DL frameworks, and is gaining popularity for its ease of use and its capability to construct the dynamic computational graph, which is now widely used by the academic community. In contrast, TensorFlow is widely used by companies, startups, and business firms to automate things and develop new systems. It has distributed training support, scalable production options, and support for mobile devices. 
Currently, the AI community has made huge efforts to develop open-source on-device frameworks like TensorFlow Lite (TFLite), Caffe2, Caffe, NCNN, and ONNX. As an on-device DL platform, TensorFlow Lite (TFLite) is the most popular framework for DL models on smartphones, as it has GPU support and is optimized for mobile devices~\cite{xu2019first,huang2021robustness}. 


\paragraph{\textbf{TFLite Models}}
TFLite models have powerful features for running models on edge devices but they do not provide APIs to access the gradient or intermediate outputs like other TensorFlow or PyTorch models. TensorFlow provides a \textit{TensorFlow Lite Converter} to convert a TensorFlow model into a TensorFlow Lite model. 
In addition, the models trained by other DL frameworks can also be converted to the TFLite model. For example, PyTorch provides the API to save the model as ONNX format, and then convert the ONNX model to TensorFlow and TFLite model \href{https://github.com/onnx/onnx-tensorflow}{Onnx-tf tool}.
For parsing the model structure and weights from the \texttt{.tflite} file, we can use the schema file \footnote{\href{https://github.com/tensorflow/tensorflow/blob/master/tensorflow/lite/schema/schema.fbs}{schema file (The link is too long to display)}} of TFLite to parse FlatBuffers and get the JSON file that contains detailed information of the \texttt{.tflite} file.

\paragraph{\textbf{ONNX Models}}
Open Neural Network Exchange (ONNX) is an open format built to represent machine learning models. ONNX defines a common file format to enable developers to use models with a variety of frameworks and tools. 
ONNX platform has various tools to support the exchange between common neural network model formats (\eg TensorFlow, PyTorch) and the ONNX model format.
For instance, the \emph{tf2onnx} tool can transform the TensorFlow model to the ONNX model accurately.
The \emph{onnx2tf} and \emph{onnx2pytorch} transform the ONNX model to corresponding TensorFlow and PyTorch models, respectively.
The \emph{onnx2tf} tool only generates a low-level saved model, which can just use the forward inference APIs (\ie the generated model is not debuggable).
Differently, the mechanism of the \emph{onnx2pytorch} is based on a rule list.
which defines the relationship between ONNX operators and PyTorch operators. It will first create a model instance by translating the ONNX model based on the rule list and generate a forward function to define the data flow at runtime. 
The converted PyTorch model is debuggable. 
Our proposed reverse engineering method based on the ONNX platform will process the non-debuggable components on the unified ONNX level and then convert them to debuggable format. So, it can be applied to multiple on-device formats such as TFLite, ONNX Runtime, and Caffe. In contrast, other transformation pipelines may not easily handle multiple on-device formats and need ad-hoc manners to build different rules for different on-device model formats.

\subsection{Adversarial Attacks for DL models}
Adversarial attacks add perturbation that can be considered a special noise to the original image to fool the DL models.
Adversarial attacks can be categorized into white-box attacks such as gradient-based attacks~\cite{croce2020reliable,goodfellow6572explaining,kurakin2016adversarial,papernot2016limitations,moosavi2016deepfool,madry2018towards,moosavi2017universal}, and black-box attacks~\cite{chen2017zoo,ilyas2018black,ilyas2018prior,guo2019simple,brendel2017decision,cheng2018query,chen2020hopskipjumpattack,mopuri2018ask}. Gradient descent (GD) is an iterative optimization algorithm, used to find a local minimum/maximum of a given function. This method is commonly used in training DL models. For the gradient-based (white-box) attack~\cite{goodfellow6572explaining,kurakin2016adversarial}, they use the gradient to compute the perturbation that can increase the model loss.
Query-based black-box attacks~\cite{brendel2017decision, chen2019hopskipjumpattack} estimate the gradients to compute the perturbation, \eg randomly update the perturbation to estimate the right update direction.
White-box attacks have full access to the model structure and its parameters to enable gradient computing. In black-box attacks, only partial information (\ie model output) about the model is available. Goodfellow \etal~\cite{goodfellow6572explaining} show that adversarial examples generated by surrogate models~\cite{papernot2017practical,zhou2020dast} can fool the target model. Therefore, for adversarial attacks on devices, Huang \etal~\cite{huang2021robustness} and Cao \etal~\cite{cao2021towards} evaluated the mobile model robustness by generating attacks from surrogate models. However, they heavily rely on the similarity between surrogate models and target models. According to this, we propose the \toolname to transform the TFLite model to the PyTorch model, to explore the security issue of model deployment.

\section{Preliminary Study}
\label{sec:preliminary}
Recall that the ONNX platform has provided various tools to support the exchange from neural network models to ONNX models. 
In this preliminary study, we would like to investigate whether these tools can be leveraged to transform a TFLite model (will be regarded as a Tensorflow model) into a PyTorch model (i.e., TFLite model $\xrightarrow[]{tf2onnx}$ ONNX model $\xrightarrow[]{onnx2pytorch}$ PyTorch model).


\subsection{Harvesting On-device Models}
\label{subsc:ondevice_models}
To conduct this preliminary study, we need to first collect a set of TFLite models that are actually included in Android apps.
Since there is no existing dataset containing a set of apps with TFLite models, we have to construct such a dataset from scratch.
To this end, we resort to the AndroZoo dataset~\cite{li2017androzoo++} to first collect a set of real-world Android apps.
AndroZoo is by far the largest app set well maintained by researchers and has been widely leveraged by researchers to support various Android-related studies.
At the moment, AndroZoo contains over 19 million Android apps. It is extremely time-consuming for us to download and scan all of them to locate TFLite models.
For the sake of simplicity, we only focus on the latest Google Play apps 
(published from 2021 to 2022)
to fulfill our study.
In total, we have collected 173,905 apps (it takes more than one month). After disassembling the apps, we find 674 of them contain TFLite-related packages (i.e., \emph{org.tensorflow}). 
All of these apps are regarded as candidates to extract TFLite models.
Among 674 apps, we were eventually able to extract 244 TFLite models, which are then taken into account to fulfill our preliminary study.

\subsection{Model Transformation Study}
\label{sec:model-study}

\begin{figure}[ht]
  \begin{center}
    \includegraphics[width=0.75\linewidth]{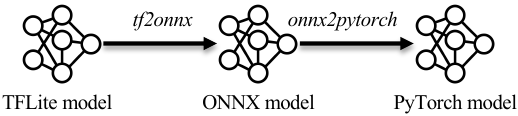}
  \end{center}
  \caption{The naive transformation flow from compiled on-device models to debuggable models in our preliminary study.}
  \label{fig:preliminary}
\end{figure}

Figure~\ref{fig:preliminary} illustrates the working process of our preliminary model transformation study.
We would like to check if existing tools can be leveraged to transform TFLite models into debuggable PyTorch models.
Since ONNX does not directly provide the tool for TFLite models, we naively regard them as Tensorflow models to fulfill this study, which is made up of three steps, as highlighted in Figure~\ref{fig:preliminary}.

\begin{itemize}[leftmargin=*]
\item\emph{\textbf{Step-1: Extractor}}
First, we extract TFLite models from Android apps by applying the well-known \emph{apktool}~\footnote{\href{https://ibotpeaches.github.io/Apktool/}{https://ibotpeaches.github.io/Apktool/}} tool to decompile the apps. \emph{Apktool} is one of the most popular tools proposed for reverse engineering Android APKs. After decompilation, we search for \texttt{.tflite} files in the decompiled folder.
\item\emph{\textbf{Step-2: tf2onnx}}
After obtaining the TFLite model, we then transform it to the ONNX model.
The advantage of achieving the transformation based on the ONNX platform is that:
(1) The ONNX model is intended to be easily modified. Adding or removing a layer from an ONNX model requires just one line of code.
(2) Our transformation tool is easy to be applied to other on-device formats like caffe2 and ONNX Runtime. 
Specifically, once the on-device caffe2 models are converted to ONNX model,
the proposed tool can be applied to convert the ONNX model to a debuggable version.

\item\emph{\textbf{Step-3: onnx2pytorch}}
After we convert the TFLite model to an ONNX model, we then convert the ONNX model to the debuggable DL model, \ie PyTorch model. Here we choose the PyTorch model as the transformation target because the API library of PyTorch is more stable than that of TensorFlow. In addition, due to its flexibility, it is simpler for us to assemble the parsed information into the debuggable PyTorch model.
\end{itemize}

\begin{figure}[t]
  \begin{center}
    \includegraphics[width=0.43\textwidth]{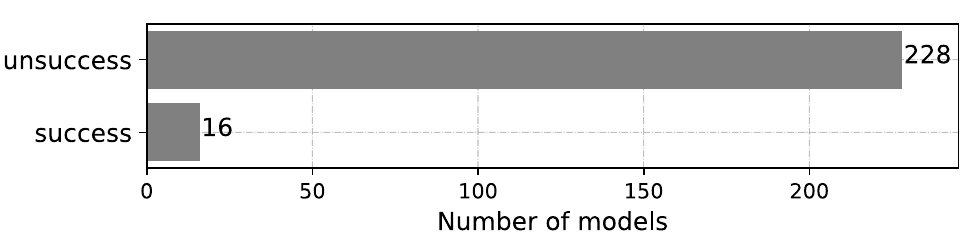}
  \end{center}
  \caption{Results of the naive transformation flow.}
  \label{fig:preliminary_result}
\end{figure}

\textbf{Experiment results}:
In this work, we use all the 244 apps (i.e., 244 TFLite models) identified in Section~\ref{subsc:ondevice_models} to fulfill this study. 
Figure~\ref{fig:preliminary_result} summarizes the experimental results.
Among the 244 models, only 16 of them can be successfully transformed into PyTorch models.
This preliminary approach yields a failure rate of 93.4\%, making it impossible to be adopted in practice to achieve our purpose, i.e., automatically transforming TFLite models to debuggable ones.

\begin{table} [tb]
  \caption{Errors types in the failure cases of the preliminary study. Note that one operator may cause multiple errors. }
  \label{tb:error}
  \footnotesize
  \centering
  \begin{tabular}{l|c|c|c}
    \hline
     \textbf{Errors}  &\textbf{Reasons} &\textbf{Count} &\textbf{Related Operators}   \\
    \hline
    \multirow{2}{*}{\scriptsize{$\texttt{Compatibility}$}}   &\multirow{2}{*}{Structure Mismatch} &\multirow{2}{*}{156}  &Quantization     \\
    &             &        &Transformation   \\
    \hline
    &\multirow{3}{*}{Operators Mismatch} &\multirow{3}{*}{100}  &  Quantization  \\

    \scriptsize{$\qquad \ \texttt{Not}$} &   &   &  Data Processing  \\
    \scriptsize{$\texttt{Implemented}$} &   &   &  Computing  \\
    \cline{2-4}
    &  \multirow{2}{*}{Operators Not Supported}  & \multirow{2}{*}{24} &   Customized    \\
    &                 &                     &   Deprecated    \\  
    \hline
    \multirow{2}{*}{\scriptsize{$\texttt{Input Type}$}}  
    &\multirow{2}{*}{Specification Mismatch} &\multirow{2}{*}{18} &  Transformation       \\
    &          &              &  Computing       \\
    \hline
  \end{tabular}
\end{table}

\textbf{Error types}:
We then go one step further to check why some models can be successful while majorities cannot.
The preliminary transformation approach fails with three types of errors: (1) {\tt Compatibility Error} (156), (2) {\tt Not Implemented Error} (124), and (3) {\tt Input Type Error} (18). {\tt Compatibility Error} appears when the model structure is not compatible with the debuggable model format. For the {\tt Not Implemented Error}, it appears when the ONNX model has operators that are not in the transformation rule list. For the {\tt Input Type Error}, it appears when the \textit{onnx2pytorch} assigns wrong inputs and parameters to the layer. The detailed analysis can be found in Section~\ref{sec:approach}.

Overall, our preliminary study shows that existing tools cannot achieve the purpose of transforming TFLite models into debuggable models.
We, therefore, argue that there is a strong need to invent new approaches to address this challenge.
Motivated by this evidence, we design and implement in this work a prototype tool called \toolname, which aims at transforming on-device models by resolving the aforementioned errors.

\begin{tcolorbox}[colback=gray!5!white,colframe=gray!85]
The fact that less than 10\% of TFLite models can be automatically transformed to debuggable PyTorch models by existing approaches. It shows that there is a strong need to invent new approaches to achieve the purpose.
\end{tcolorbox}

\section{Approach}
\label{sec:approach}

We now detail our approach proposed to transform on-device TFLite models into debuggable PyTorch models.
\begin{figure*}[tb]
  \begin{center}
    \includegraphics[width=0.85\textwidth]{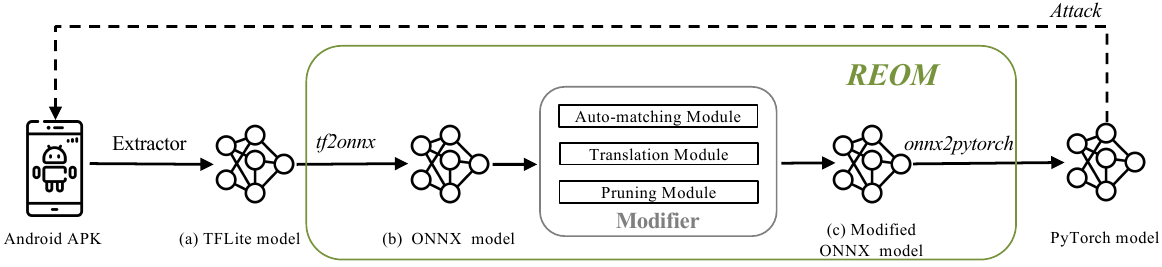}
  \end{center}
  \vspace{-0.3em}
  \caption{The overview working process of \toolname. We use the (a), (b), and (c) to define the state of the model in \toolname.
  }
  \label{fig:workflow}
\end{figure*}
Before presenting our method,
we first analyze the errors (see Table \ref{tb:error}) in existing tools: 
\begin{itemize}[leftmargin=*]
    \item {\small\textbf{\texttt{Compatibility - Structure Mismatch:}}} 
    To accelerate the computation on mobile devices equipped with mobile CPU and mobile GPU, compiled on-device models are optimized to contain some different data types and model structures with debuggable models.
    For example, \texttt{float32} will be converted to \texttt{uint8} when compiling the debuggable model to the on-device model. 
    When converting the on-device model back to ONNX models, some extra operators (\eg quantization operators, transformation operators) will be created to make it compatible with the ONNX data types and structures.
    Since the extra structures are non-debuggable, the structure mismatch will unfortunately result in the failure of the transformation. 
    \textbf{In our approach, we propose the \emph{Pruning Module} to resolve this problem.}
    
    \item  {\small\textbf{\texttt{Not Implemented - Operator Mismatch:}} }
    Some compiled operators of on-device models are optimized for mobile computing.
    The optimized operators transformed from TFLite model are not compatible with the debuggable model format. 
    For example, 
    when the data type of the on-device operator is \texttt{uint8} (optimized data type for on-device model), this operator will not be debuggable because DL frameworks like PyTorch and TensorFlow do not have debuggable API for \texttt{uint8}.
    In addition, TFLite defines many unique operators (\ie mismatched operators for debuggable models) supported by its corresponding library.
    However, such unique operators are not supported by other DL frameworks (\eg PyTorch, TensorFlow). 
    \textbf{Therefore, in this work, we propose the \emph{Translation Module} to bridge the mismatched operators.}

    \item  {\small\textbf{\texttt{Not Implemented - Operators Not Supported:}}}
    Those customized \textbf{operators are not supported} by other DL models, resulting in TFLite cannot be directly transformed.
    For example, if developers want to implement an advanced Convolutional operator in their model but this advanced function is not supported by the current version of the on-device DL framework, they could add their customized C/C++ implementation of the advanced function to their TFLite model. Besides, developers can name this customized operator and define the interface freely. So, other DL frameworks cannot identify this customized operator because it is not included in the operator library.
    This problem also exists on the deprecated operators of TFLite models.
    Fortunately, on-device operators are usually a subset of the debuggable operators' library. 
    \textbf{Therefore, we propose \emph{Auto-matching Module} to identify the equivalent operators in the debuggable model format to fulfill the transformation.}

    \item {\small\textbf{\texttt{Input Type - Specification Mismatch:}}}
    The specifications of some operators (\eg computing operators, transformation operators) may vary in different model formats, such as the order of inputs and the range of parameters.
    Therefore, the setting of the converted debuggable operators may be wrong. \textbf{Because these errors need to be resolved in an ad-hoc manner, we omit them in this paper.}
    
\end{itemize}

Therefore, we design and implement in this work a prototype tool called \toolname, which aims at removing the aforementioned errors while maximizing the similarity between the on-device model and the converted debuggable model. Figure~\ref{fig:workflow} presents the overall working process of our proposed \toolname, which essentially contains four steps to achieve its purpose, \ie transforming a TFLite model into a debuggable PyTorch model for security exploitation. The four steps are \ding{172} Extractor, \ding{173} \emph{tf2onnx}, \ding{174} Modifier, and \ding{175} \emph{onnx2pytorch}. The details of Extractor and \emph{tf2onnx} can be found in Section \ref{sec:model-study}.

\subsection{Modifier}
\label{subsec:modifier}

We now detail the Modifier of \toolname with the three modules to resolve the aforementioned problems, respectively.

\begin{figure}[tb]
  \begin{center}
    \includegraphics[width=0.43\textwidth]{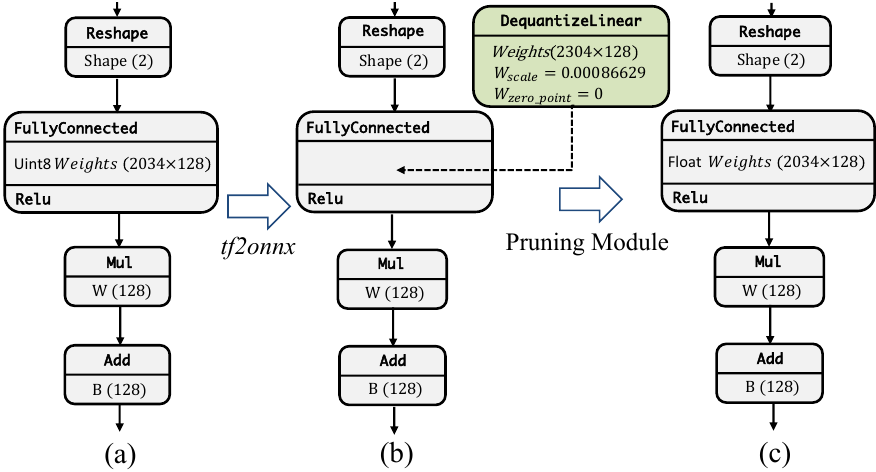}
  \end{center}
  \caption{Demonstration of the structure mismatch. (a) The TFLite model. (b)The TFLite-converted ONNX model. (c) The ONNX model modified by our Pruning Module.
  }
  \label{fig:structure_mismatch}
\end{figure}

\textbf{Pruning Module}:
To solve the structure mismatch issue, we propose Pruning Module.
Before presenting technique details, we first show an example of the structure mismatch in Figure \ref{fig:structure_mismatch}. 
When the on-device model (Figure~\ref{fig:structure_mismatch}(a)) tries to convert to the debuggable format, some non-debuggable components are not compatible with the debuggable format. As shown in green areas of Figure~\ref{fig:structure_mismatch}(b), those non-debuggable components need to be processed before connecting with other debuggable components. Therefore, the ``extra'' part will be produced, but it will confuse debuggable DL libraries because they do not consider this special case in many functions like gradient computing.
For example, in Figure~\ref{fig:structure_mismatch},
the type of weights in \texttt{FullyConnected} layer is \texttt{uint8}, which cannot be handled by ONNX operators and debuggable model format. To address it, the \texttt{uint8} tensor needs to be transformed by the formula: $y = (x - y_0) \times y' $, where the $y_0$ and $y'$ are stored in the model files and are the zero-point and scale parameters of this \texttt{uint8} tensor. After converting to ONNX format, it needs to attach an extra new operator \texttt{DequantizeLinear} to achieve the above transformation.  

Unfortunately, the TFLite-converted ONNX model with such an extra branch is still not compatible with the debuggable model format like PyTorch. 
To address this problem, the Pruning Module is proposed to correct the mismatched structure so that it will be compatible with the debuggable format.
To remove the mismatched structure, we first find the suspect non-debuggable extra operators using pruning rules. 
Specifically, we analyze the ONNX operator library to identify the operators that can be used to transform the model weights (\eg DequantizeLinear, Reshape, Transpose). These operators are potential extra operators. The complete pruning rules list can be found in our code repository. 
If an operator conforms to our definition of extra operators (\ie the operator is in the pruning list, uses the fixed tensor data as its input, and produces output data as the next operator's weights), we will remove the extra operator, and compute the corresponding transformed weights for debuggable models, \eg using the transformation formula of the extra operator $y = (x - y_0) \times y' $, which is shown in Figure~\ref{fig:structure_mismatch}(c), to remove the non-debuggable branch.

\textbf{Translation Module}: 
Translation Module is used to solve the operator mismatch issue. Different DL libraries have different specifications (\eg interface, parameters, and algorithm principle) for the equivalent operators, especially for the on-device DL library and debuggable library. 
Some on-device operators (\eg \texttt{QuantizeLinear}) is not compatible with the debuggable format because DL libraries do not provide support for those on-device-related operators in their debuggable platform.
So we cannot directly map the operator (\ie mismatched operators) from the on-device form to the debuggable version like existing tools to achieve our purpose.

To address this problem, the Translation Module translates the mismatched operators (\eg {\tt SpaceToDepth, QuantizeLinear} in TFLite$\rightarrow$ONNX$\rightarrow$PyTorch) into several basic operators that are supported by debuggable formats. 
For example, the formula of {\tt QuantizeLinear} in TFLite and ONNX can be presented as:
\begin{align}
   y = \delta(\frac{x}{y'} + y_0) 
\end{align}
where the $y_0$ and $y'$ are the zero-point and scale parameters of the operator, respectively.
Note that, the $x/y'$ is float division.
The $\delta$ is a saturation parameter that saturates the value to [0, 255], and then converts its data type to \texttt{uint8}. 
To make it debuggable and compatible with PyTorch, the \texttt{QuantizeLinear} operator can be divided into two separate operators \texttt{Add} (matrix addition) and \texttt{Div} (matrix division).
Those two operators support the computation of float values.
The formula can be shown as:
\begin{align}
    y = \texttt{Clip}(\texttt{Add}(\texttt{Div}(x, y'), y_0)),
\end{align}
where the \texttt{Clip} clip all elements of input into the range [0, 255] when the target data type is \texttt{uint8}.
To automatically translate the mismatched operators, we analyze the operator list of the DL platform where we want to process the issue (\eg ONNX in our study), and identify which operator does not have the matched debuggable operators. Then, we create a translation list that defines the mapping from mismatched operator list $L_o^T$ to the debuggable equivalence list $\widehat{L_o^T}$. Note that we build translation rules for all operators in the translation list. If the Translation Module finds an operator that is in the translation list, it will replace the mismatch operator with the equivalent operator combination. 

\begin{figure}[tb]
  \begin{center}
    \includegraphics[width=0.4\textwidth]{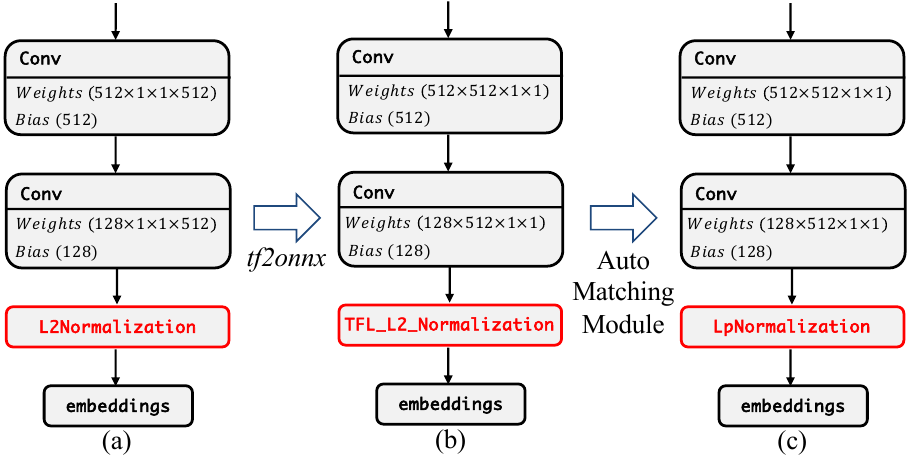}
  \end{center}
  \vspace{-0.3em}
  \caption{Demonstration of non-supported operators. (a) The TFLite model. (b) The TFLite-converted ONNX model. (c) The ONNX model modified by our Auto-matching Module.}
  \label{fig:operator_not_support}
\end{figure}

\textbf{Auto-matching Module}:
Auto-matching Module is used to handle the customized or deprecated operators that are not supported by other debuggable model formats. This is because mobile DL developers sometimes use customized implementations on DL libraries to achieve their purpose. Sometimes they don't use the latest version of DL libraries to build their model, some operators are deprecated in the latest versions when we try to transform them. To enhance the exception-handling ability of our method, we introduce the Auto-matching Module.
The example (\texttt{TFL\_L2\_NORMALIZATION}) is shown in Figure \ref{fig:operator_not_support}.
Unfortunately, PyTorch does not support the \texttt{TFL\_L2\_NORMALIZATION}, hence does not support the transformation of this operator.

\begin{table}[t]
	\centering
        \small
	\begin{tabular}{lll}
		\toprule
            $\bf Algorithm~1 $ Auto-matching \\ 
		\midrule

		\textbf{Input}:$ \text{ supported debuggable operators list } L, \text{ input operator } o_x, $ \\
		$  \text{similarity threshold } \alpha, \text{random inputs } \mathbf{I} $ \\
		
		\textbf{Output}: syntactic matched debuggable operator $o_y$\\
		$1: \textbf{If } \ o_x \text{ not in } L:$ \\
		$2: \qquad \textbf{For } i \text{ in range(lenth(L))}:  $ \\
		$3: \qquad \qquad \text{compute the } D_{ro}^i \text{ by } o_x \text{ and } L[i] \text{ using Equation 3}  $ \\
		$4: \qquad \textbf{end For} $  \\
		$5: \qquad \text{Sort } \mathbf{D_{ro}}\!=\![D_{ro}^1, \dots, D_{ro}^{lenth(L)} ]  $ \text{in descending order}  \\
            $6: \qquad \textbf{For } D_{ro}^y \text{ in } \mathbf{D_{ro}}: $   \\
		$7: \qquad \qquad \textbf{If   } ||f_{o_x}(\mathbf{I}) - f_{o_y}(\mathbf{I})||_2 \le \alpha:  $   \\
		$8: \qquad \qquad \qquad \textbf{return } o_y  $   \\

		\bottomrule	
	\end{tabular}
	\label{matching}
	\vspace{-0.5em}
\end{table}
To this end, we propose a three-step syntactic matching approach to find an equivalent supported operator to replace the non-supported operator, which is shown in Algorithm 1.

First, when our proposed tool~\toolname finds the non-supported operator (ONNX operators in our tool), which is not in the operator list of other DL model formats, the Auto-matching Module will compute the distance between the supported debuggable operators list $L$ and the non-supported operator.
The similarity between the \texttt{op\_types} of the non-supported operator and the supported debuggable operator can be obtained as follows:

\begin{align}
    D_{r o}=\frac{2 K_{m}}{\left|S_{1}\right|+\left|S_{2}\right|}
\end{align}
where the $D_{r o}$ is the similarity metric and $0 \leq D_{\text {ro }} \leq 1$. The $S_1$ and $S_2$ are the keyword string of the non-supported operator and the supported debuggable operator, respectively. $K_m$ is the number of matched characters.

Second, we rank the supported operators list $L$ as the distance between the non-supported operator and the supported operator. Then we can find the most similar supported operator $o_y$ with the non-supported operator.

Third, the Auto-matching Module will replace the unsupported operators with the most similar supported operator $o_y$. Then, it calculates the function similarity between the unsupported operators and $o_y$ by comparing the output difference of original on-device models (TFLite model in our study) and the modified model (ONNX model in our study) with the same inputs $\mathbf{I}$ (the number of inputs is 100 in our experiments). If the $l_2$ difference $||f_{o_x}(\mathbf{I}) - f_{o_y}(\mathbf{I})||_2$ between the non-supported operator and $o_y$ is smaller than a threshold $\alpha$ (we set it to 0.1 in default), the $o_y$ is the matched operator.

For example, as shown in the (b)$\to$(c) of Figure \ref{fig:operator_not_support}, the customized (non-supported) operator of the TFLite-converted ONNX model ({\tt TFL\_L2\_NORMALIZATION} in Figure \ref{fig:operator_not_support} (b)) is converted to equivalent ONNX operator ({\tt LpNormalization} in Figure \ref{fig:operator_not_support} (c)) with the Auto-matching Module. 

\subsection{Converting to The Debuggable Model}
After modifying the converted ONNX model, it will save the modified model as the new {\tt.onnx} file.
Then, we use the \textit{onnx2pytorch} tool to load the structure information and parameter of the ONNX model and assemble them into the Python code using PyTorch API. 
It is also worth noting here that the generated debuggable model will share the same structure and parameters as the on-device model extracted from real-world Android apps.
Consequently, the two models should share the same capabilities.
They should also share the same attacking surfaces.
In other words, the attack scenarios applicable to the debuggable PyTorch model could be directly applied to attack the on-device TFLite model (indicated via dotted line in Figure~\ref{fig:workflow}). 
We present the experimental results in Section 5.

\section{Evaluation}
\label{sec:eva}

Towards checking if our research objective is achieved, 
we propose to answer the following three key research questions.

\begin{itemize}

\item \textbf{RQ1:} How effective is our approach in achieving automated model transformation?

\item \textbf{RQ2:} How accurate is the transformation approach?

\item \textbf{RQ3:} Can on-device models be directly attacked via \toolname-based white-box strategies?

\end{itemize}

\noindent
\textbf{Dataset Construction}
In the evaluation section, we use the same dataset construction strategy in Section \ref{subsc:ondevice_models} and answer the research question using \toolname.

\begin{table} [tb]
  \caption{Transformation performance of the proposed \toolname. }
  \label{tb:convert_performance}
  \small
  \centering
  \begin{tabular}{lcccc}
    \toprule
    Error Types      &  Reasons       & Count & Success & Fail   \\
    \hline
    \footnotesize{$\texttt{Compatibility}$} & Structure Mismatch      & 156    & 156     & 0        \\
    \cline{1-5}
    \multirow{2}{*}{\footnotesize{$\texttt{Not\!Implemented}$}} &  Operator Mismatch   & 100   & 100    & 0       \\

    &  Operator Not Supported & 24    & 6     & 18   \\
    \cline{1-5}
    \footnotesize{$\texttt{Input Type}$} &  Specification Mismatch & 18    & 18    & 0 \\
    \hline
  \end{tabular}
\end{table}

\begin{table} [t]
  \caption{Comparison between different $\alpha$ values of Algorithm 1. }
  \label{tb:alpha}
  \small
  \centering
  \begin{tabular}{lccccc}
    \toprule
    \small{Operator Not Supported}  & $0$  & $0.001$ & $0.01$ & $0.1$ & $100$ \\
    \hline
    Success Cases  & 0  &  0   &  2  & 6  & 24  \\

    Fail Cases     & 24 &  24   &  22  &  18 & 0   \\

    \hline
  \end{tabular}
\end{table}

\subsection{RQ1: Effectiveness}

In this part, we use all 244 apps (\ie 244 TFLite models) to fulfill our study. 
\textbf{As shown in Figure~\ref{fig:preliminary_result}, among the 244 models, only 16 of them can be successfully transformed into PyTorch models by existing tools.
This baseline approach yields a failure rate of 93.4\%}, making it impossible to be adopted in practice to achieve our purpose, \ie automatically transforming TFLite models to debuggable ones.

In contrast, \textbf{\toolname is able to successfully transform 226 of them, giving a success rate of 92.6\%}. Note that the 16 models that can be handled by existing tools do not have the non-debuggable component. For those models, the debuggable models produced by our method are the same as the models generated by existing tools.
Table \ref{tb:convert_performance} further breaks down the detailed results with respect to the three types of issues summarized previously.
Note that a given TFLite model may encounter several errors.
Hence, the total number of errors (\ie 289) is slightly larger than the number of TFLite models.
All the failure cases are caused by the Operator Not Supported issue, for which the Auto-matching Module cannot find an existing ONNX operator that is equivalent to the customized or deprecated TFLite operator.

Observant readers may have noticed that our approach has taken the parameter $\alpha$ to determine whether the newly generated ONNX operator (because of non-supported TFLite operators) should be accepted in the Auto-matching Module.
We now go one step deeper to evaluate the sensitivity of this parameter.
As shown in Table~\ref{tb:alpha}, when $\alpha$ is set to be 0.1 (the default value), 6 of 24 models with custom non-supported operators can be successfully converted (18 failures).
When decreasing this threshold, the failure rate will increase.
In the worst case, when $\alpha$  is set to zero, none of the non-supported operator problems can be resolved.
However, in another extreme setting, when setting the $\alpha$ to be 100, all the non-supported operators can be resolved, \ie mapped to newly generated operators that are accepted by the debuggable model format.
Subsequently, all the TFlite models can be successfully converted. However, such transformation will not make much sense as the transformed models may not perform the same as their source models.
In this work, the default $\alpha$ value 0.1 is set based on our empirical experience, under which the output difference between the original TFLite model and its PyTorch counterpart can be controlled within a distance of 0.1.

\begin{figure}[t]
  \begin{center}
    \includegraphics[width=0.4\textwidth]{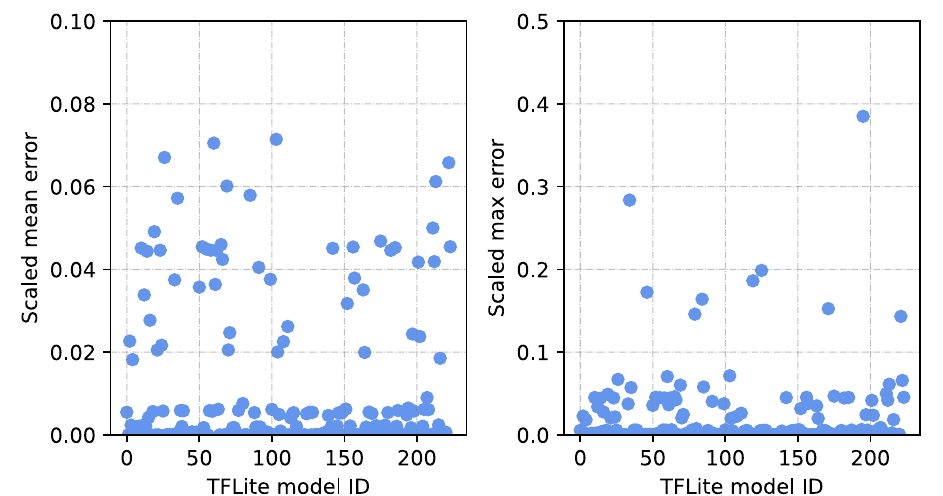}
  \end{center}
  \caption{The scaled difference for models from TFLite model to the converted model. The x-axis refers to the ID of the on-device models. Here we plot the transformation difference of all collected models.}
  \label{fig:error_sta}
\end{figure}

\begin{tcolorbox}[colback=gray!5!white,colframe=gray!85]
 \textbf{Answer to RQ1:} The proposed \toolname can successfully transform over 90\% of TFLite models to debuggable models.
  Two out of the three modifier modules have achieved 100\% correctness, while our best-effort attempts implemented in the remaining module have also been demonstrated to be useful.
\end{tcolorbox}

\begin{table} [tb]
  \caption{Demonstration of which operator of the TFLite model will affect the transformation accuracy. We use the (min, max) to define the typical difference range, where the min and max are the minimal and maximal transformation differences for the TFLite model with the specific operator.}
  \label{tb:operation_affect}
  \small
  \centering
  \begin{tabular}{lcc}
    \toprule
    Category                          & Operator        &  Difference Range                \\
    \hline
    \multirow{2}{*}{Computing Difference} & \texttt{DequantizeLinear}  &   \multirow{2}{*}{(0.001, 0.01)}  \\
                                         & \texttt{QuantizeLinear}    &     \\
    \cline{1-3}
    \multirow{2}{*}{API Difference}           & \texttt{Resize}           &   \multirow{2}{*}{(0.001, 0.05)}   \\
                                         & \texttt{Upsample}         &                                      \\

    \hline
  \end{tabular}
\end{table}

\begin{table*} [t]
  \caption{Classification accuracy of on-device models and converted debuggable models on test images. 
  Each app has a different test dataset. The dataset list can be found in the shared code repository. The TFLite models are collected from the work~\cite{huang2021robustness} and the TensorFlow Hub.
  }
  \vspace{+0.5em}
  \label{tb:cls_acc}
  \centering
  \begin{tabular}{lc|ccccccccc}
  
    \hline
    & Models & Fruit  & Skin cancer & object & Sign language & Plant & Cassava disease & Plant disease &  Insect  &  Bird                                                  \\
    \hline

    \multirow{3}{*}{Accuracy }
                            & \toolname  & 100.00\%   & 80.51\%   & 70.59\%  & 98.71\% & 95.00\%  & 91.76\% & 93.20\% & 96.48\%  & 92.07\%     \\

                            & Source model                  & 100.00\%   & 80.60\%   & 70.62\%    & 98.24\% & 95.08\% & 93.12\% & 93.20\% & 96.48\%  & 92.50\%  \\
    \cline{2-11}
                  &   Difference                & 0.00\%   & 0.09\%   & 0.03\%    & 0.47\% & 0.08\% & 1.36\% & 0.00\% & 0.00\%  & 0.43\%  \\
    \hline
  \end{tabular}
\end{table*}

\subsection{RQ2: Accuracy}

We then compare the output similarity between the transformed models and source models to evaluate the accuracy. 
Given a pair of models (\ie a TFLite model and its PyTorch counterpart), with the same inputs, the accuracy of our approach is evaluated based on the similarity of the outputs yielded by the two models.
The similar the results are, the higher the accuracy will be.
In practice, we use the collected 244 on-device models as the test set. We generate 100 random inputs as the specification of the TFLite model and compare the outputs between the TFLite models and their debuggable PyTorch versions. However, the output ranges of the on-device models are different. To standardize the output difference between the two models, we use the scaled mean transformation difference, which can be calculated as follows:
\begin{equation}
\begin{aligned}
  d = \frac{1}{rk} \sum_{i=1}^k |\vec{y_i} - \hat{\vec{y_i}} |,
\end{aligned}    
\end{equation}
where $d$ represents the difference between the PyTorch model and the TFLite model.
The $\vec{y_i}$ and $\hat{\vec{y_i}}$ are the outputs of the TFLite model and the converted PyTorch model, respectively. $k$ is the element number of the $\vec{y_i}$. It means we calculate the average difference for each output data point. For example, if the output is an image, the $d$ is the average difference for each pixel. $r$ is the range of the source on-device model's output. For example, if the data type of output data is {\tt Uint8}, the output range $
r$ is $255-0 = 255$. However, we cannot know the actual output range of the TFLite model when the data type is {\tt Float32}. To estimate the output range, we use the $r = \text{max}(\hat{\vec{y_i}}) - \text{min}(\hat{\vec{y_i}})$ as the output range, where the \text{max} and \text{min} are the functions to compute the maximal and minimal value of a vector of matrix, respectively. Therefore, \textbf{the estimated difference in our experiments may be lower than the actual difference because the output range in our calculation is smaller than the actual value}. Similarly, to compute the scaled maximal transformation difference, the formula is shown as follows:
\begin{equation}
\begin{aligned}
  d = \frac{1}{r} \text{max}( |\vec{y} - \hat{\vec{y}} |).
\end{aligned}    
\end{equation}

The result is highlighted in Figure \ref{fig:error_sta}.
The transformed PyTorch model generally has a very small difference compared with the TFLite model. \textbf{Most cases have a difference of less than 0.001. Some cases have a difference from 0.001 to 0.08}, which is also small. 
For the cases where the transformation errors are larger than 0.1, the output debuggable model may have a large difference from the source model. It means attackers cannot achieve the white-box attacks based on our method for these cases. However, it will still outperform the black-box attacks used in the existing attacking evaluation studies~\cite{huang2021robustness,huang2022smart}.
In addition, our method will not affect the overall accuracy of models (\cf the accuracy difference of converted models and source models in Table~\ref{tb:cls_acc}). 
\textbf{It demonstrates that the converted debuggable models have very similar accuracy to the source on-device models.} The difference exists because the on-device models usually have 8-bit precision but PyTorch debuggable formats only support 16 or 32-bit precision.

Besides, we analyze which operator of the TFLite model will affect the transformation difference. In Table \ref{tb:operation_affect}.
We find two main reasons that can affect the transformation difference. One is the computational difference. Another one is the API difference. For the computational difference, the converted PyTorch model runs on the {\tt float32} data type. However, when the TFLite model has some operators like {\tt DequantizeLinear} and {\tt QuantizeLinear}, it will cause the computational difference between the TFLite model and the converted PyTorch model. For the API difference, some TFLite APIs and PyTorch APIs are fundamentally different. For example, the {\tt Resize} API of TFLite and PyTorch will use a basic {\tt Interpolate} operation. It determines how to compute the value of resized tensors. However, TFLite has more methods to implement the {\tt Interpolate} operation. If the {\tt Interpolate} operation of TFLite layers is not supported by PyTorch, \toolname will use a substitute {\tt Interpolate} operation to execute the {\tt Resize}. It will cause the output of PyTorch models to be slightly different from the output of TensorFlow models.

\begin{tcolorbox}[colback=gray!5!white,colframe=gray!85]
\textbf{Answer to RQ2:} The proposed \toolname approach can achieve high accuracy of the transformation. The performances of the generated PyTorch models are generally very similar to their original on-device TFLite models,
which enables security exploitation in the white-box setting.
\end{tcolorbox}

\begin{figure*}[tb]
  \begin{center}
    \includegraphics[width=0.78\textwidth]{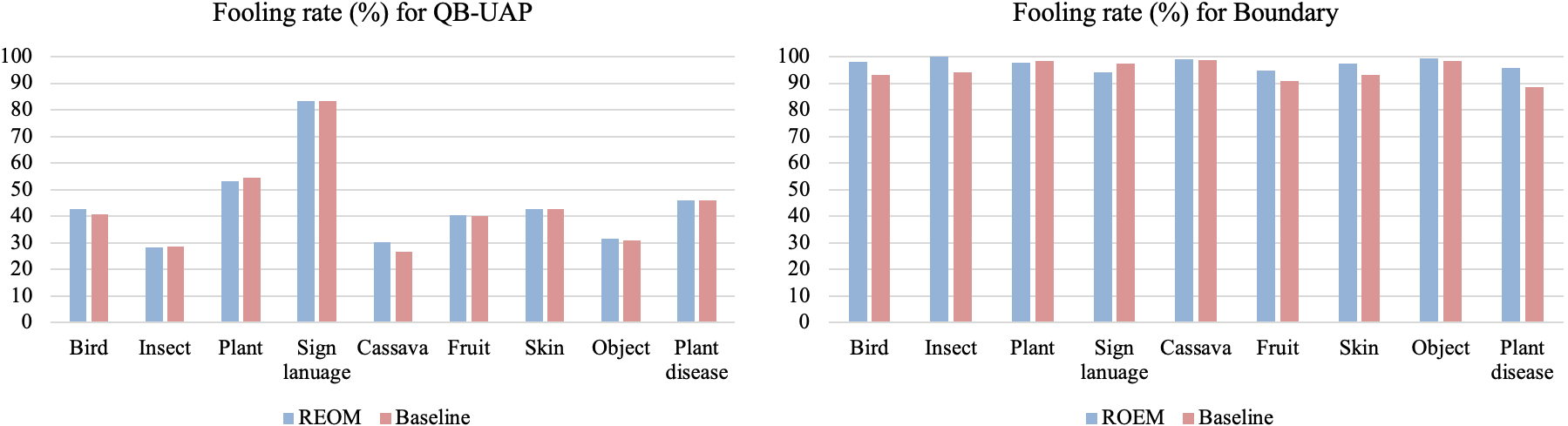}
  \end{center}
  \caption{The Fooling rate of the debuggable model converted by \toolname and the source on-device models using black-box attack methods. It shows that converted models have a similar fooling rate on black-box attacks with source models.}
  \label{fig:black_att}
\end{figure*}

\begin{table*} [tb]
  \caption{Fooling rate (\%) of non-targeted and targeted attacks using white-box attack methods. `$\ell_2$': the $\ell_2$ distance (perturbation magnitude). `BIM': Basic Iterative Method~\cite{kurakin2016adversarial}. `PGD': Projected Gradient Descent for generating attacks~\cite{madry2018towards}. `\toolname': we generate attacks on the converted PyTorch model and transfer the adversarial example to attack the target on-device model. 
  `Baseline': we collect the pre-trained model like the works~\cite{huang2021robustness,huang2022smart} and fine-tune the model on our collected dataset like the work~\cite{cao2021towards}, then transfer the attacks to attack the target on-device model.}
  \vspace{+0.5em}
  \resizebox{.95\linewidth}{!}{%
  \label{tb:attack}
  \centering
  \begin{tabular}{lc|ccccccccc}
  
    \toprule
    \multicolumn{11}{c}{\textbf{Non-targeted Attack}}                                                                  \\
     \hline
    & & Fruit  & Skin cancer & ImageNet & Sign language & Plant & Cassava disease & Plant disease &  Insect  &  Bird                                                  \\
    \hline
    $\ell_2$                  & Attack                 & BIM$\mid$PGD                         & BIM$\mid$PGD       & BIM$\mid$PGD  & BIM$\mid$PGD & BIM$\mid$PGD & BIM$\mid$PGD & BIM$\mid$PGD  & BIM$\mid$PGD  & BIM$\mid$PGD    \\
    \hline
    \multirow{2}{*}{0.01}
                            & \toolname  & 2.90$\mid$2.90   & 1.91$\mid$1.91   & 89.77$\mid$89.03  & 0.60$\mid$0.60 & 72.51$\mid$72.31 & 57.89$\mid$57.97 & 0.75$\mid$0.80 & 41.61$\mid$41.82  & 56.08$\mid$56.00     \\

                            & Baseline                  & 0$\mid$0   & 0$\mid$0   & 4.40$\mid$4.56    & 0$\mid$0 & 2.67$\mid$2.64 & 1.18$\mid$1.31 & 0$\mid$0 & 0.78$\mid$0.62  & 1.32$\mid$1.30  \\
    \hline
    \multirow{2}{*}{0.1}
                            & \toolname            & 28.12$\mid$28.12   & 20.70$\mid$20.58   & 96.14$\mid$96.14   & 31.89$\mid$31.05 & 80.53$\mid$80.40 & 67.20$\mid$67.11 & 14.75$\mid$14.79 & 53.73$\mid$53.77  & 68.37$\mid$67.99   \\

                            & Baseline                  & 0.22$\mid$0.22   & 0.31$\mid$0.31   & 5.73$\mid$5.62    & 2.98$\mid$2.74 & 19.74$\mid$19.59 & 8.37$\mid$8.37 & 0.36$\mid$0.36 & 1.82$\mid$1.86 & 4.51$\mid$4.51   \\

    \hline
    \multirow{2}{*}{1.0}
                            & \toolname            & 99.78$\mid$99.55   & 100.00$\mid$100.00   & 99.65$\mid$99.65   & 100.00$\mid$100.00 & 99.96$\mid$99.92 & 99.87$\mid$99.87 & 100.00$\mid$100.00 & 96.31$\mid$96.27 &99.49$\mid$99.49    \\

                            & Baseline                  & 3.79$\mid$4.02   & 2.67$\mid$3.02   & 10.78$\mid$10.23   & 12.93$\mid$12.42 & 41.42$\mid$41.40 & 30.82$\mid$30.86 & 1.96$\mid$1.96 & 24.32$\mid$23.89 & 27.68 $\mid$ 27.72   \\

    \hline
    \multicolumn{11}{c}{\textbf{Targeted Attack}}                                                                  \\
     \hline
   $\ell_2$                  & Attack                 & BIM$\mid$PGD                         & BIM$\mid$PGD       & BIM$\mid$PGD  & BIM$\mid$PGD & BIM$\mid$PGD & BIM$\mid$PGD & BIM$\mid$PGD  & BIM$\mid$PGD  & BIM$\mid$PGD    \\
    \hline
    \multirow{3}{*}{0.01}
                            & \toolname  & 2.90$\mid$2.90   & 1.36$\mid$1.36  & 3.97$\mid$3.97  & 0.04$\mid$0.08 & 0.86$\mid$0.90 & 19.09$\mid$19.09 & 0.12$\mid$0.12 & 0.27$\mid$0.27  & 1.02$\mid$0.90    \\

                            & Baseline                  & 0$\mid$0   & 0$\mid$0   & 0$\mid$0    & 0$\mid$0 & 0$\mid$0 & 0$\mid$0 & 0$\mid$0 & 0$\mid$0  & 0$\mid$0  \\
    \hline
    \multirow{2}{*}{0.1}
                            & \toolname            & 1.61$\mid$1.61   & 10.79$\mid$10.79   & 14.94$\mid$15.93   & 3.63$\mid$3.63 & 3.63$\mid$3.67 & 28.08$\mid$28.12 & 1.09$\mid$1.09 & 1.06$\mid$1.13  & 3.64$\mid$3.71   \\
                            
                            & Baseline                  & 0$\mid$0   & 0.76$\mid$0.76   & 0$\mid$0    & 0$\mid$0 & 0$\mid$0 & 
                            2.04$\mid$2.11 & 0$\mid$0 & 0$\mid$0 & 0$\mid$0   \\

    \hline
    \multirow{2}{*}{1.0}
                            & \toolname            & 86.18$\mid$86.41   & 99.88$\mid$99.88   & 98.21$\mid$98.41   & 89.74$\mid$89.74 & 96.09$\mid$95.90 & 92.56$\mid$92.61 & 90.12$\mid$90.34 & 72.13$\mid$71.89 & 96.17$\mid$96.13    \\
                            
                            & Baseline                  & 1.32$\mid$1.37   & 4.13$\mid$4.15   & 0.02$\mid$0.02   & 2.95$\mid$2.95 & 3.72$\mid$3.68 & 10.54$\mid$10.59 & 1.27$\mid$1.23 & 2.94$\mid$2.88 & 0.64$\mid$0.68    \\      
    \hline
  \end{tabular}}
\end{table*}

\subsection{RQ3: Supporting White-box Attacks}

We evaluate the attacking performance of \toolname to check whether attackers can directly perform white-box attacks for on-device models. 
We choose nine TFLite classification models of the work~\cite{huang2021robustness} and the TensorFlow Hub to answer this research question. We choose these models because we can find large-scale public datasets (see our code repository) to evaluate the attack success rate to show the effectiveness. For the fruit app, we identify 848 images whose categories correspond to the task scope of models and then use these images as the test set. For other apps, we randomly sample 10000 images from the large-scale datasets as test sets. 


Then, we evaluate the attacking performance of the on-device model using the proposed \toolname. In~\cite{huang2021robustness,cao2021towards}, they focus on generating adversarial attacks by the surrogate model to mislead the target on-device model in the black-box setting. The performance of adversarial attacks generated by the surrogate model relies on the similarity between the target model and the surrogate model. Our study proposes a method for transforming the compiled TFLite model into a debuggable model, eliminating the need to search for or train a surrogate model in order to achieve white-box-like attack performance. Therefore, we will go in-depth into the on-device adversarial attacks and show how our tool can be a general method to evaluate the robustness of the on-device model.

We calculate the attack success rate (\ie fooling rate) by $p = \frac{n}{m}$, where $n$ and $m$ are the number of successful adversarial examples and the number of images that can be correctly classified by the model, respectively. Note that we only perform attacks on the image which is correctly classified by the target model. $n$ is the number of successful adversarial examples. For \textbf{non-targeted attacks}, the attack succeeds when the target model outputs the wrong labels for the inputs. For \textbf{targeted attacks}, the attack succeeds when the target model outputs a specific wrong label. Generally, targeted attacks are more difficult to produce than non-targeted attacks.


We first demonstrate how is the similarity of the converted model and the source on-device model (\ie baseline) on black-box attacks, which is shown in Figure~\ref{fig:black_att}. We use two different black-box attack methods, one individual method Boundary \cite{brendel2017decision} that needs to generate different attack perturbations for each input and one universal method QB-UAP~\cite{wu2020decision} that produces a universal perturbation for all inputs. They do not need gradient information of the attacked model to evaluate the robustness of models. Here we set the $\ell_2$ attack distance to 15. The hyper-parameters of the attack method in our experiments are the same as the parameters in the original paper. \textbf{We find the converted model has a similar black-box attack performance to the source on-device model.}

\begin{figure*}[tb]
  \begin{center}
    \includegraphics[width=0.82\textwidth]{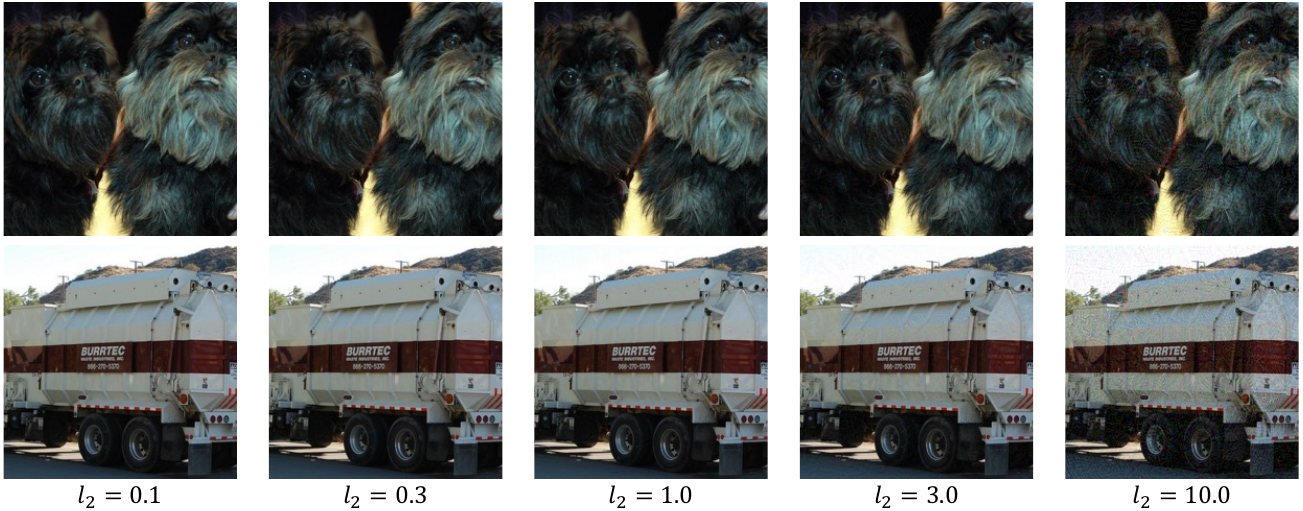}
  \end{center}
  \caption{The visualization of different perturbation distances for the image classification model. }
  \label{fig:visual}
   \vspace{-0.5em}
\end{figure*}

Then, we use the method proposed by Huang~\etal~\cite{huang2021robustness} to collect similar pre-trained models with the target mobile models from the TensorFlow Hub based on the structure and weights similarity, and then fine-tuned them on the training set as the surrogate model. We choose the attacks generated by the surrogate model as the baseline.  
For \toolname, we transform the TFLite model into the PyTorch model as the surrogate model. Then, we compare the fooling rate between our method and the baseline. We use two well-known white-box attack methods, BIM~\cite{kurakin2016adversarial} and PGD~\cite{madry2018towards}, to generate the attack. They are the most common methods used in the robustness evaluation for DL models~\cite{zhou2022adversarial}. For these methods, we set the number of iterations to 400. The step size of each iteration is set to 0.0001, 0.001, and 0.04 for the perturbation distances $\ell_2$ are 0.01, 0.1, and 1.0, respectively.

As the results are shown in Table \ref{tb:attack}, if we use the proposed tool to get the converted PyTorch model as the surrogate, the attack performance will significantly increase compared with that leveraging conventional transfer attacks. 
The debuggable models can indeed support stronger attacks. Compared with previous methods of generating attacks using surrogate models, \textbf{attackers can achieve higher attack success rates (10.23\%→89.03\% in ImageNet apps) with a hundred times smaller attack perturbations (1.0→0.01) based on the proposed \toolname framework.} The visualization of different perturbation distances is shown in Figure~\ref{fig:visual}. REOM-based attacks can achieve high attack performances using small perturbations that are imperceptible to humans.



\begin{tcolorbox}[colback=gray!5!white,colframe=gray!85]
\textbf{Answer to RQ3:} The proposed \toolname approach is indeed useful for helping security analysts to evaluate the security of on-device TFLite models. Experimental results demonstrate the converted model can be considered as the debuggable version of the source model for security exploitation. On-device models can indeed be directly attacked via REOM-based white-box strategies.
\end{tcolorbox}

\begin{figure}[h!]
  \begin{center}
    \includegraphics[width=0.4\textwidth]{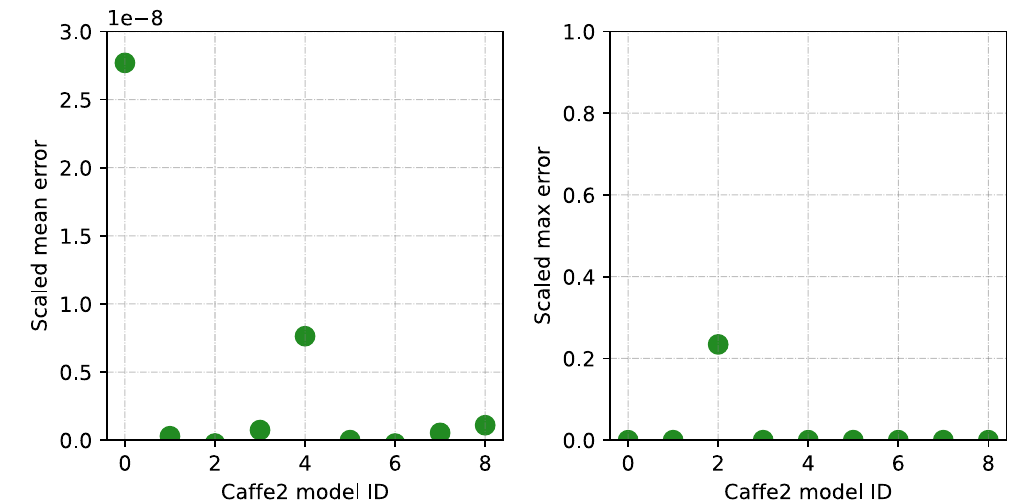}
  \end{center}
  \caption{The scaled difference between the Caffe2 models and the converted debuggable model. Our tool can transform 9 of 10 Caffe2 models. The x-axis refers to the ID (0-8) of the Caffe2 models. }
  \label{fig:error_sta_caffe}
\end{figure}

\section{Discussion}
In this section, we will discuss the genericity, other properties of our method, and potential defense strategies.
\label{sec:discussion}


\textbf{Generalizability of Our Approach:}
Although our method is designed for the most commonly used on-device model format TFLite, our approach should also work for other formats because the \emph{Modifier} handles the non-debuggable component in the ONNX level, which has a unified model representation. 
we believe our approach should also work for transforming other on-device formats like Caffe2 to the debuggable model format. 
To experimentally validate this, we collect 10 Caffe2 models from the Caffe2 model zoo \cite{caffe2zoo}
to evaluate the effectiveness of \toolname on the other on-device format. 
By default, only two out of the ten models can be successfully handled by the existing toolchain.
We then integrate our approach into the process by applying our \emph{Modifier} to automatically modify the intermediate ONNX model generated by the built-in conversion tool of \emph{Caffe2}.
Now, 9 of 10 \emph{Caffe2} models can be transformed into PyTorch models. All the converted models have a scaled difference less than $3\times10^-{8}$, as shown in Figure~\ref{fig:error_sta_caffe}.
It shows \toolname is indeed generic and can handle multiple on-device formats. 
However, if the on-device model cannot be converted to the ONNX model, our method does not work. 
As for our future work, we commit to evaluating and enhancing the generalizability of our approach with more on-device models.

\begin{figure}[tb]
  \begin{center}
    \includegraphics[width=0.43\textwidth]{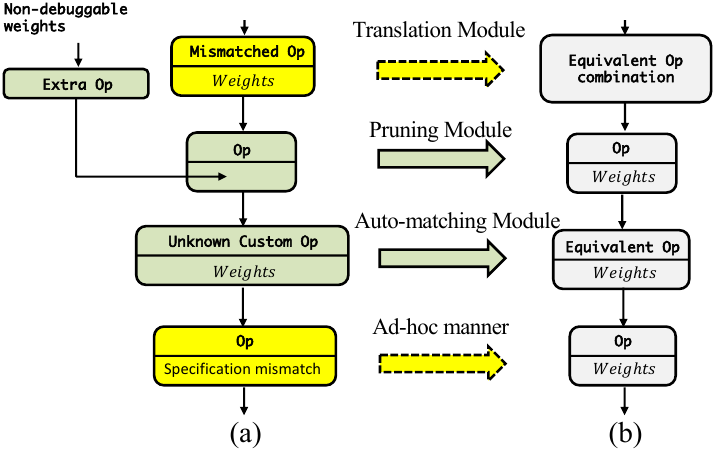}
  \end{center}
  \caption{The meta-model of how our \toolname solves the four problems. (a) general representation of non-debuggable models. (b) general representation of transformed debuggable models. }
  \label{fig:meta_model}
   \vspace{-0.5em}
\end{figure}

Although our method is generic, some parts (\eg translation rule list) of our approaches may need to be updated frequently to support the future versions of DL model formats or adapt to other model formats. As shown in Figure~\ref{fig:meta_model}, The green arrow means this problem exists because of fundamental differences between the non-debuggable models and debuggable models, \ie having a clear definition and can be solved by a unified solution. The dotted yellow arrow means this problem exists because of occasional differences between the non-debuggable models and debuggable models, \eg the solution may be various in different versions of DL libraries. For example, some operators are debuggable in this version but may be non-debuggable in the next version. 

In addition, we use adversarial attacks to evaluate the effectiveness of our method. The process of producing other kinds of attacks~\cite{shokri2017membership,fredrikson2015model} based on the proposed \toolname is similar. Attackers need to first get the debuggable model from the on-device model and then generate other attacks. However, the attack generation for different kinds of attacks may be different. For example, model inversion attacks~\cite{fredrikson2015model} use the debuggable model to find the input that can produce the same output as the source on-device model.

\textbf{On Increasing the Attack Surface of on-device models:}
Existing studies~\cite{huang2022smart, li2021deeppayload, sang2023beyond} for evaluating the robustness of on-device models cannot access them as white-box ones. We conduct this study to explore the real risks of DL models on devices. 
It indicates attackers can fully access the on-device model through reverse engineering for most real-world Apps.
Our experimental results demonstrate that the converted debuggable version of on-device models can indeed have a similar prediction performance compared with the original model.
This result strongly supports our hypothesis that it is indeed possible to conduct direct white-box attacks for target on-device models. More importantly, the Direct white-box method using \toolname can significantly increase the attacking performance, and achieve higher attack success rates (10.23\%$\to$89.03\%) with a hundred times smaller attack perturbations (1.0$\to$0.01). 
So, existing studies~\cite{huang2022smart, li2021deeppayload, sang2023beyond} for evaluating the robustness of on-device models usually miss the fact that attackers can bridge the gap by reverse engineering. Our paper leverages empirical software engineering methods to reveal the real risk of on-device models, which is underestimated by the existing studies.

\textbf{Enabling the white-box testing:}
Our study enables direct testing on deployed DL models like TFLite models.
Although there are many white-box testing methods to evaluate the DL models~\cite{ma2018deepgauge,pei2017deepxplore}, these methods are designed for debuggable DL models. However, the deployed model may not be debuggable (or differentiable), such as the TFLite model. Existing white-box testing strategies, which are more efficient than black-box ones, cannot be directly applied to the on-device models.
Therefore, our contribution lies in enabling direct white-box testing of compiled DL models.


\textbf{Potential defense strategies:}
The observation from the evaluation of our method shows reverse engineering the on-device model relies on the effectiveness of transformation rules. 
At the moment, The proposed REOM can successfully transform over 90\% of TFLite models into debuggable models as almost all on-device operators (226 cases) can be reverse-engineered into debuggable ones.
There are indeed some corner cases for which \toolname fails because of customized operators (18 cases).
Except for this, we believe there might be more options for defending against direct white-box attacks~\cite{liu2022deep}.
One possible approach developers could consider is to split models, e.g., by defining multiple sub-models in training and then compiling them into different model files. Attackers need to understand the source code (the code in apps is usually obfuscated) to know how to assemble them. However, this method may increase the inference time of on-device models because it needs to load and parse the model twice.
Another approach could be to implement model obfuscation~\cite{zhou2023modelobfuscator}, e.g., by replacing the keyword of the \texttt{conv2d} operator to a random string,  and build a compatible customized TFLite library. However, this method may increase memory and time consumption because it needs to parse the obfuscated information before computing the output using the model information.


\section{THREATS TO VALIDITY}

We now discuss the potential threats to the validity of this work.

First, our proposed conversion tool \toolname is based on the ONNX platform, and we evaluate its performance on TFLite models and caffe2 models. However, some on-device model formats may have a higher level of security (\eg do not use high-level representations like TVM models), which may disable the model parsing based on the operator-to-operator transformation rule list, including our approach proposed in this work. In this case, those on-device models are safe. However, reverse engineering methods may overcome this problem by modifying the conversion rules, \eg building a mapping list from ONNX operators to TVM model representations. 

Second, the development of the DL library is in rapid change. It may affect the performance of reverse engineering when the library updates the model format or conducts other major evolutionary changes. In such a case, the reverse engineering tool may fail to convert on-device models to debuggable versions.
Therefore, we argue that there is a strong need for our approach to be aware of the evolution of given DL frameworks.

\section{Conclusion}
This study evaluates the importance of developing a reverse engineering tool that can transform the TFLite model into the debuggable PyTorch model. 
Such transformation can enable attackers to perform direct white-box attacks for evaluating the vulnerability of on-device models.
To achieve this, we propose a \toolname framework to transform the on-device model into the PyTorch model. Our proposed \toolname has three steps: (1) first, we use the \textit{tf2onnx} tool to convert the TFLite to the ONNX model. (2) Second, we propose a three-module modifier, which has Pruning Module, Translation Module, and Auto-matching Module. It can modify the ONNX model to make it compatible with the debuggable PyTorch format. (3) Finally, the modified ONNX model can be successfully transformed into the PyTorch model by the \textit{onnx2pytorch} tool. 
Experiments show the \toolname can effectively transform most TFLite models to PyTorch models, with small transformation differences compared with the original TFLite model. Then, we test our method on adversarial attacks and find that 
on-device models can be directly attacked via white-box
strategies. The current model deployment strategy is at serious risk.
it enables attackers to perform white-box attacks on on-device models.
In future works, we will comprehensively analyze the security and privacy issues of on-device models using the proposed \toolname.

\section{Data Availability}
Our codes and experimental data are available online (\url{https://github.com/zhoumingyi/REOM}), which includes the setup steps and how to reproduce the main results of this paper.

\begin{acks}
This work is partially supported by the Open Foundation of Yunnan Key Laboratory of Software Engineering under Grant No.2023SE102, by the National Natural Science Foundation of China under Grant No.62202026 and No.62172214, and by Guangxi Collaborative Innovation Center of Multi-source Information Integration and Intelligent Processing.
\end{acks}






\bibliographystyle{ACM-Reference-Format}
\bibliography{acmart}

\end{document}